\documentclass[11pt]{amsart}

\usepackage{graphicx}  
\usepackage{amsmath}  
\usepackage{amssymb,amsthm}
\usepackage{color}
\usepackage{tikz}

\newcommand{\Z}{{\mathbb Z}}

\let\epsilon\varepsilon
\let\theta\vartheta

\newtheorem{theorem}{Theorem}
\newtheorem{lemma}[theorem]{Lemma}

\newtheorem{remark}[theorem]{Remark}

\title[Limit of a diatomic FPU system]{\bf The monoatomic FPU system \\as a limit of a diatomic FPU system}

\author{Dmitry E. Pelinovsky}
\address[D. Pelinovsky]{Department of Mathematics and Statistics, McMaster University,
Hamilton, Ontario, Canada, L8S 4K1}
\email{dmpeli@math.mcmaster.ca}

\author{Guido Schneider}
\address[G. Schneider]{Institut f\"{u}r Analysis, Dynamik und Modellierung,
Universit\"{a}t Stuttgart, Pfaffenwaldring 57, D-70569 Stuttgart, Germany}
\email{guido.schneider@mathematik.uni-stuttgart.de}

\date{}                                           

\begin{document}
\maketitle

\begin{abstract}
We consider a diatomic infinite Fermi--Pasta--Ulam (FPU) system with light and heavy particles.
For a small mass ratio, we prove  error estimates for the approximation
of the dynamics of this system by the dynamics of the monoatomic FPU system.
The light particles are squeezed by the heavy particles at the mean value of their displacements.
The error estimates are derived by means of the energy method
and hold for sufficiently long times, for which the dynamics of the monoatomic FPU system is observed.
The approximation result is restricted to sufficiently small displacements of the heavy particles
relatively to each other.
\end{abstract}

\section{Introduction}

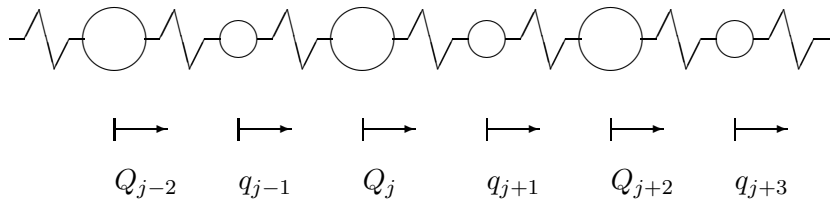
\begin{figure}[htbp] 
   \centering
    \setlength{\unitlength}{1cm}
   \begin{picture}(12, 4)


 \put(0.5,3){\line(1,0){0.2}}
  \put(0.7,3){\line(1,2){0.2}}
   \put(0.9,3.4){\line(1,-4){0.2}}
  \put(1.1,2.6){\line(1,2){0.2}}
\put(1.3,3){\line(1,0){0.2}}

\put(2.3,3){\line(1,0){0.2}}
  \put(2.5,3){\line(1,2){0.2}}
   \put(2.7,3.4){\line(1,-4){0.2}}
  \put(2.9,2.6){\line(1,2){0.2}}
\put(3.1,3){\line(1,0){0.2}}

\put(3.8,3){\line(1,0){0.2}}
  \put(4,3){\line(1,2){0.2}}
   \put(4.2,3.4){\line(1,-4){0.2}}
  \put(4.4,2.6){\line(1,2){0.2}}
\put(4.6,3){\line(1,0){0.2}}

\put(5.6,3){\line(1,0){0.2}}
  \put(5.8,3){\line(1,2){0.2}}
   \put(6,3.4){\line(1,-4){0.2}}
  \put(6.2,2.6){\line(1,2){0.2}}
\put(6.4,3){\line(1,0){0.2}}

\put(7.1,3){\line(1,0){0.2}}
  \put(7.3,3){\line(1,2){0.2}}
   \put(7.5,3.4){\line(1,-4){0.2}}
  \put(7.7,2.6){\line(1,2){0.2}}
\put(7.9,3){\line(1,0){0.2}}

\put(8.9,3){\line(1,0){0.2}}
  \put(9.1,3){\line(1,2){0.2}}
   \put(9.3,3.4){\line(1,-4){0.2}}
  \put(9.5,2.6){\line(1,2){0.2}}
\put(9.7,3){\line(1,0){0.2}}

\put(10.4,3){\line(1,0){0.2}}
  \put(10.6,3){\line(1,2){0.2}}
   \put(10.8,3.4){\line(1,-4){0.2}}
  \put(11,2.6){\line(1,2){0.2}}
\put(11.2,3){\line(1,0){0.2}}

 \put(1.9,3){\circle{0.8}}
 \put(3.55,3){\circle{0.5}}

 \put(5.2,3){\circle{0.8}}
 \put(6.85,3){\circle{0.5}}

 \put(8.5,3){\circle{0.8}}
 \put(10.15,3){\circle{0.5}}

 \put(1.9,1){$Q_{j-2}$}
  \put(3.55,1){$q_{j-1}$}
  \put(5.2,1){$Q_{j}$}
  \put(6.85,1){$q_{j+1}$}
    \put(8.5,1){$Q_{j+2}$}
  \put(10.15,1){$q_{j+3}$}

     \put(1.9,1.65){\line(0,1){0.3}}
   \put(1.9,1.8){\vector(1,0){0.7}}
     \put(3.55,1.65){\line(0,1){0.3}}
   \put(3.55,1.8){\vector(1,0){0.7}}
  \put(5.2,1.65){\line(0,1){0.3}}
   \put(5.2,1.8){\vector(1,0){0.7}}
     \put(6.85,1.65){\line(0,1){0.3}}
   \put(6.85,1.8){\vector(1,0){0.7}}
  \put(8.5,1.65){\line(0,1){0.3}}
   \put(8.5,1.8){\vector(1,0){0.7}}
  \put(10.15,1.65){\line(0,1){0.3}}
   \put(10.15,1.8){\vector(1,0){0.7}}
\end{picture}
   \caption{A diatomic FPU system with heavy and light particles.}
   \label{fig1}
\end{figure}

We consider a diatomic infinite Fermi--Pasta--Ulam (FPU) system depicted schematically on
Figure \ref{fig1}. Displacements of heavy particles are denoted by $ Q_j $ with $j \in 2 \mathbb{Z}$,
whereas displacements of light particles are denoted by $ q_j $, with $ j \in 2 \mathbb{Z} + 1$.
For convenience, we normalize the mass of the heavy particles to unity and denote the mass ratio
between masses of light and heavy particles by the parameter $ \varepsilon^2 $. The total
energy of the diatomic system is
\begin{equation}
\label{energy-FPU}
E = \sum_{j \in 2 \mathbb{Z}} \frac{1}{2} \dot{Q}_j^2 + \frac{1}{2} \varepsilon^2 \dot{q}_{j+1}^2 +
W(q_{j+1}-Q_j) + W(Q_j - q_{j-1}),
\end{equation}
where the dot denotes the derivative in time $t$ and $W : \mathbb{R} \mapsto \mathbb{R}$ is a smooth
potential for the pairwise interaction force between the adjacent light and heavy particles.
Equations of motion are generated from the total energy (\ref{energy-FPU}) by using the standard
symplectic structure for the dynamics of particles. They are written in the form:
\begin{eqnarray}
\ddot{Q}_{j} & = & W'(q_{j+1}-Q_j) - W'(Q_j-q_{j-1}) \label{merit2}, \\
\varepsilon^2 \ddot{q}_{j+1} & = & W'(Q_{j+2}-q_{j+1}) - W'(q_{j+1}-Q_{j}) \label{merit3},
\end{eqnarray}
where $ j \in 2 \mathbb{Z} $.

The dynamics of diatomic lattices, e.g. propagation of traveling solitary waves, has always been important
in physical applications and has been studied in numerous works, e.g. \cite{ph1,ph2,ph3}.
More recently, such diatomic systems were considered in the context of granular chains \cite{chain2,chain3,chain1}.
In particular, the authors of \cite{chain2} proposed to consider the following reduction
of the diatomic system in the limit of vanishing mass ratio $ \varepsilon \to 0 $:
$$
0 = W'(Q_{j+2}-q_{j+1}) - W'(q_{j+1}-Q_{j}) \quad \Rightarrow \quad Q_{j+2}-q_{j+1}= q_{j+1}-Q_{j},
$$
which yields
\begin{equation} \label{merit0}
q_{j+1} = \frac{Q_{j+2} + Q_{j}}{2}.
\end{equation}
If $q_{j+1}$ is expressed by \eqref{merit0}, the dynamics of the heavy particles
is governed by the monoatomic FPU system:
\begin{equation} \label{merit1}
\ddot{Q}_{j}  =  W'\left( \frac{Q_{j+2} - Q_{j}}{2}\right) - W'\left( \frac{Q_{j} - Q_{j-2}}{2} \right),
\end{equation}
where $j \in 2 \mathbb{Z}$. It follows from (\ref{merit0}) and (\ref{merit1}) that the light particles
are squeezed by the heavy particles and move according to the mean values of their displacements,
whereas the heavy particles move according to their pairwise interactions.

Numerical results on existence and non-existence of traveling solitary waves in the diatomic system (\ref{merit3})
which are close to the traveling solitary waves of the monoatomic system (\ref{merit1}) were reported in \cite{chain2}.
These numerical results inspired a number of analytical works where the authors developed the existence theory
for traveling solitary waves with oscillatory tails \cite{FW18,HW17}, beyond-all-order theory \cite{L20,LP18},
and the linearized analysis of perturbations \cite{Vain16}. {\em It is the purpose of this paper to give
rigorous error estimates for this approximation in the context of the initial-value problem.}

Note that the small mass ratio limit for diatomic FPU system has been considered before in the context of
the existence of breathers \cite{JN04,JN08,LSM97,Yoshimura11} and traveling periodic waves \cite{BP13,J12,Qi14,Qi15}.
However, these works rely on the ideas of the so-called anti-continuum limit, for which the heavy particles do not move
after rescaling of the time variable, whereas the light particles perform uncoupled oscillations
in between the heavy particles. The limit (\ref{merit0}) and (\ref{merit1}) is clearly different from the anti-continuum limit.

Other relevant results on travelling solitary waves in diatomic lattices include
persistence results near the equal mass ratio limit \cite{FH20},
asymptotic approximations near the long-wave limit \cite{GMWZ14,W19}, and
numerically assisted study of radiation generated from long-wave solitons
in the time evolution \cite{GSWW19}.

We shall now present the main approximation theorem. We use the standard notation
$\ell^2$ to denote square summable sequences equipped with the norm
$$
\| u \|_{\ell^2} := \left( \sum_{k \in \mathbb{Z}} |u_k|^2 \right)^{1/2},
$$
from which it is obvious that $\sup_{k \in \mathbb{Z}} |u_k| \leq \| u \|_{\ell^2}$.
Another useful property of the $\ell^2$ space is being a Banach algebra with respect to pointwise
multiplication.

\begin{theorem} \label{thmain}
Assume that $ Q^* \in C^1([0,T_0],\ell^2) $ is  a solution of the scalar FPU lattice \eqref{merit1}
wit $W \in C^3(\mathbb{R})$ and $W''(0) > 0$ for a fixed $T_0 > 0$.
There exist $ \varepsilon_0 > 0 $, $ C_0 > 0$, and $ C > 0 $ such that for all $ \varepsilon \in (0,\varepsilon_0) $,
the following is true. If $(Q(0),q(0)) \in \ell^2 \times \ell^2$ satisfy the bound
\begin{equation}
\label{initial-bound}
\sup_{j \in 2 \mathbb{Z}} \left(|Q_{j}(0) - Q^*_{j}(0)| +
\left|q_{j+1}(0) - \frac{Q^*_{j+2} (0)+ Q^*_{j}(0)}{2} \right| \right) \leq \varepsilon,
\end{equation}
and $ Q^* \in C^1([0,T_0],\ell^2) $ satisfy the bound
\begin{equation}
\label{relative-displacement}
\sup_{t \in [0,T_0]} \sup_{j \in 2 \Z } |Q^*_{j+2}(t) -Q^*_j(t)| < C_0,
\end{equation}
then there exists the unique solution $ (Q,q) \in C^1([0,T_0],\ell^2 \times \ell^2 ) $
to the diatomic FPU system \eqref{merit2}-\eqref{merit3}, which satisfies the bound
\begin{equation}
\label{final-bound}
\sup_{t \in [0,T_0]} \sup_{j \in 2 \mathbb{Z}} \left( |Q_{j}(t) - Q^*_{j}(t)| +
\left| q_{j+1}(t) - \frac{Q^*_{j+2} (t)+ Q^*_{j}(t)}{2} \right| \right) \leq C \varepsilon .
\end{equation}
\end{theorem}

\begin{remark}{\rm
The approximation result of Theorem \ref{thmain}
is nontrivial since the right hand side of the associated first order system to \eqref{merit2}, and \eqref{merit3} multiplied with $ \varepsilon^{-2} $, is of order  $\mathcal{O}(\varepsilon^{-1}) $.
Standard Gronwall's inequality only gives estimates on an $\mathcal{O}(\varepsilon) $-time scale
and not on the natural  $\mathcal{O}(1) $-time scale.}
\end{remark}

\begin{remark}{\rm
Approximation results for systems with a small perturbation parameter in front of one time derivative,
similar to system \eqref{merit2}-\eqref{merit3} have been considered in \cite{BSSZ19}. However, the abstract theorem
from  \cite{BSSZ19} does not apply since the nonlinear interaction appearing here is different as the
one considered in Eq. (14) of \cite{BSSZ19}.  The approach in \cite{BSSZ19} is based on a normal form transformation,
where the proof presented here is based on a suitable choice of coordinates and energy estimates.}
\end{remark}

The remainder of the paper is organized as follows.
In Section \ref{secneu2}, we rewrite the diatomic FPU system in new coordinates which are
more suitable to express perturbations to the motion given by the limit system \eqref{merit0} and \eqref{merit1}.
The bounds in Theorem \ref{thmain} are obtained with the energy estimates in Section  \ref{secneu4}
for the simple case with $W'(u) =  u +  u^2$. Generalizations to other nonlinear interaction potentials
$W(u)$ are discussed in Section \ref{secdisc}.

\medskip

{\bf Acknowledgement.}
G. Schneider is partially supported by the Deutsche Forschungsgemeinschaft DFG through the CRC 1173 ``Wave phenomena".
D. Pelinovsky is partially supported by the grant of the President of the Russian Federation for
scientific research of leading scientific schools of the Russian Federation NSh-2485.2020.5.

\section{Change of coordinates}
\label{secneu2}

By using suitable chosen coordinates, we will separate the fast and slow dynamics of the diatomic FPU system
\eqref{merit2}--(\ref{merit3}) and will introduce perturbations to the  motion given by the
limit system \eqref{merit0}--\eqref{merit1}. Note that the same choice of coordinates was used 
in \cite{HW17} in the study of traveling waves. Let us set
$$
U_j := \frac12 (Q_{j+2}-Q_j) \quad \textrm{and} \quad w_{j+1} := q_{j+1} - \frac12 (Q_{j+2}+Q_j),
$$
so that
$$
q_{j+1}-Q_j = U_j + w_{j+1} \quad \textrm{and} \quad Q_{j+2}-q_{j+1} = U_j - w_{j+1}.
$$
The diatomic FPU system (\ref{merit2})--(\ref{merit3}) is now written as
\begin{eqnarray*}
2 \ddot{U}_{j}  
& = & W'(U_{j+2}+ w_{j+3}) - W'(U_{j}-w_{j+1})-W'(U_{j}+w_{j+1}) + W'(U_{j-2}-w_{j-1})
\end{eqnarray*}
and
\begin{eqnarray*}
\varepsilon^2 \ddot{w}_{j+1} 
& = & W'(U_{j}-w_{j+1}) - W'(U_{j}+w_{j+1}) -  \frac12 \varepsilon^2 W'(U_{j+2}+ w_{j+3})  \\
&& +  \frac12 \varepsilon^2W'(U_{j}-w_{j+1})
-  \frac12 \varepsilon^2W'(U_{j}+w_{j+1}) +  \frac12 \varepsilon^2 W'(U_{j-2}-w_{j-1}).
\end{eqnarray*}
For the particular choice $  W'(u) =  u +  u^2  $ we obtain
\begin{eqnarray*}
W'(U_{j}-w_{j+1}) + W'(U_{j}+w_{j+1}) & = & 2 U_j + 2 U_j^2 + 2 w_{j+1}^2, \\
W'(U_{j}-w_{j+1}) - W'(U_{j}+w_{j+1}) & = & - 2 w_{j+1} - 4 U_j w_{j+1},
\end{eqnarray*}
which yields the following system of equations:
\begin{eqnarray}
\label{eq-U}
\ddot{U}_{j} + U_j + U_j^2 + w_{j+1}^2 & = &  g(U_{j+2},U_{j-2},w_{j+3},w_{j-1}), \\
\label{eq-w}
\varepsilon^2 \ddot{w}_{j+1} + (2 + \varepsilon^2) w_{j+1} (1 + 2 U_j)
& = &  \varepsilon^2 h(U_{j+2},U_{j-2},w_{j+3},w_{j-1}),
\end{eqnarray}
where
\begin{eqnarray*}
g(U_{j+2},U_{j-2},w_{j+3},w_{j-1}) & = & \frac{1}{2} W'(U_{j+2} + w_{j+3}) + \frac{1}{2} W'(U_{j-2} - w_{j-1}), \\
h(U_{j+2},U_{j-2},w_{j+3},w_{j-1}) & = & -\frac{1}{2} W'(U_{j+2} + w_{j+3}) + \frac{1}{2} W'(U_{j-2} - w_{j-1}).
\end{eqnarray*}
The dynamics of $U$ and $w$ occurs now at two different scales: $U$ changes on the time scale of $\mathcal{O}(1)$,
whereas $w$ changes on the faster time scale of $\mathcal{O}(\varepsilon)$. The approximation result of Theorem \ref{thmain}
justifies the dynamics of $U$ on the time scale of $\mathcal{O}(1)$. The dynamics of $w$ is slaved to the dynamics of $U$
on this time scale.

\section{The error estimates}
\label{secneu4}

The leading-order approximation in the new coordinates is denoted by $ (U,w) = (\Psi,0) $, where $\Psi$ satisfies
\begin{equation} \label{Psieq}
\ddot{\Psi}_j + \Psi_j + \Psi_j^2 = g(\Psi_{j+2},\Psi_{j-2},0,0).
\end{equation}
After inserting this approximation into the equations of motion, the remaining terms are collected in
the residual, which is given by
 \begin{eqnarray*}
 \textrm{Res}_{U,j} & = & 0 , \\
 \textrm{Res}_{w,j} & = &  \varepsilon^2 h(\Psi_{j+2},\Psi_{j-2},0,0).
 \end{eqnarray*}
Thanks for $\ell^2$ being a Banach algebra with respect to pointwise multiplication, the residual terms obey the following estimate.

 \begin{lemma} \label{lem41}
Assume that $ \Psi \in C([0,T_0],\ell^2) $ is a solution of the scalar equation \eqref{Psieq} for some $T_0 > 0$.
Then there exists a constant $ C > 0 $ that depends on $\Psi$ such that for all $ \varepsilon \in (0,1) $ we have
\begin{eqnarray}\label{pepper0}
 \sup_{t \in [0,T_0]} \| \textrm{Res}_{w} \|_{\ell^2} \leq C \varepsilon^2.
\end{eqnarray}
 \end{lemma}

For estimating the difference between the approximation and true solutions we introduce the error functions $ R $ and $ v $ by
using the decomposition
\begin{equation}
\label{decomposition}
U_j = \Psi_j + \varepsilon R_j \quad \textrm{and} \quad w_{j+1}= \varepsilon v_{j+1}.
\end{equation}
These functions satisfy the following system
\begin{eqnarray*}
\ddot{R}_{j} + R_j + 2 \Psi_j R_j  + \varepsilon R_j^2 + \varepsilon v_{j+1}^2 & = &  L_{U,j}(\Psi)(R,v) +
\varepsilon N_{U,j}(\Psi,R,v), \\
\varepsilon^2 \ddot{v}_{j+1} + 2 v_{j+1} (1 + 2 \Psi_j + 2 \varepsilon R_j )
& = &  \varepsilon^2 L_{w,j}(\Psi)(R,v) + \varepsilon^3 N_{w,j}(\Psi,R,v) +  \varepsilon^{-1} \textrm{Res}_{w,j},
\end{eqnarray*}
where the linear terms in $(R,v)$ are given by
\begin{eqnarray*}
L_{U,j}(\Psi)(R,v) & = & \frac{1}{2} (R_{j+2} + R_{j-2}) + \frac{1}{2} (v_{j+3} - v_{j-1}) \\
& & + \Psi_{j+2} (R_{j+2} + v_{j+3}) + \Psi_{j-2} (R_{j-2} - v_{j-1}), \\
L_{w,j}(\Psi)(R,v) & = & - (1 + 2 \Psi_j) v_{j+1} - \frac{1}{2} (R_{j+2} - R_{j-2}) - \frac{1}{2} (v_{j+3} + v_{j-1}) \\
& & - \Psi_{j+2} (R_{j+2} + v_{j+3}) + \Psi_{j-2} (R_{j-2} - v_{j-1}),
\end{eqnarray*}
and quadratic terms in $(R,v)$ are given by
\begin{eqnarray*}
N_{U,j}(\Psi,R,v) & = & \frac{1}{2} (R_{j+2} + v_{j+3})^2 + \frac{1}{2} (R_{j-2} - v_{j-1})^2,\\
N_{w,j}(\Psi,R,v) & = & - 2 R_j v_{j+1} - \frac{1}{2} (R_{j+2} + v_{j+3})^2 + \frac{1}{2} (R_{j-2} - v_{j-1})^2.
\end{eqnarray*}
Thanks again for $\ell^2$ being a Banach algebra with respect to pointwise multiplication,
the linear and quadratic terms obey the following estimate.

 \begin{lemma} \label{lem42}
Assume that $ \Psi \in C([0,T_0],\ell^2) $ is a solution of the scalar equation \eqref{Psieq} for some $T_0 > 0$.
Then there exists a constant $ C > 0 $ that depends on $\Psi$ such that for all $ \varepsilon \in (0,1) $ we have
\begin{eqnarray}\label{pepper1}
\| L_{U}(\Psi)(R,v)  \|_{\ell^2} + \| L_{w}(\Psi)(R,v)  \|_{\ell^2} & \leq & C (\|  R \|_{\ell^2}+ \| v \|_{\ell^2}), \\
\| N_{U}(\Psi,R,v) \|_{\ell^2} + \| N_{w}(\Psi,R,v) \|_{\ell^2} & \leq & C  (\|  R \|_{\ell^2}^2+ \| v \|_{\ell^2}^2).
\label{pepper4}
\end{eqnarray}
 \end{lemma}

The dynamics of the error functions is estimated with the help of a suitable chosen energy.
We define the energy function by
\begin{eqnarray}
\label{energy-linear}
	E(t) = \frac{1}{2} \sum_{j \in 2 \Z} \dot{R}_j^2 + R_j^2 + \varepsilon^2 \dot{v}_{j+1}^2 + 2 v_{j+1}^2 + 2 \Psi_j (R_j^2 + 2 v_{j+1}^2)
+ 4 \varepsilon R_j v_{j+1}^2.
\end{eqnarray}
Computing the time derivative of $E(t)$ yields
\begin{eqnarray}
\label{balance-1}
\frac{d}{dt} E(t) & = & \langle \dot{R}, \ddot{R} + R + 2 \Psi R + 2 \varepsilon v^2 \rangle_{\ell^2} \\
& & + \langle \dot{v}, \varepsilon^2 \ddot{v} + 2 v + 4 \Psi v + 4 \varepsilon R v \rangle_{\ell^2}
+ \langle \dot{\Psi}, R^2 + 2 v^2 \rangle_{\ell^2},
\nonumber
\end{eqnarray}
where $(\Psi R)_j = \Psi_j R_j$ and $(\Psi v)_j = \Psi_j v_{j+1}$.
By substituting the dynamical equations for $(R,v)$ into \eqref{balance-1}, we obtain
\begin{eqnarray}
\label{balance-2}
\frac{d}{dt} E(t) & = & \langle \dot{R}, -\varepsilon R^2 + \varepsilon v^2 + L_U(\Psi)(R,v) + \varepsilon N_U(\Psi,R,v) \rangle_{\ell^2}\\
\nonumber
& & + \langle \varepsilon \dot{v}, \varepsilon L_w(\Psi)(R,v) + \varepsilon^2 N_w(\Psi,R,v)
+ \varepsilon^{-2} \textrm{Res}_w \rangle_{\ell^2}\\
& & + \langle \dot{\Psi}, R^2 + 2 v^2 \rangle_{\ell^2}.
\nonumber
\end{eqnarray}

The energy function controls the perturbations and their time derivative if displacements
of heavy particles relatively to each other are sufficiently small, as in the condition \eqref{relative-displacement}
of Theorem \ref{thmain}. The following lemma gives the corresponding result.

\begin{lemma}
\label{lem43}
Assume that $ \Psi \in C([0,T_0],\ell^2) $ is a solution of the scalar equation \eqref{Psieq} for some $T_0 > 0$.
There exist $ C_0 > 0$ and $ C > 0 $ such that if
\begin{equation}
\label{bound-on-Psi}
\sup_{t \in [0,T_0]} \sup_{j \in 2 \Z } |\Psi_j(t)| < C_0,
\end{equation}
then
\begin{equation}
\label{energy-control}
\| \dot{R}(t) \|_{\ell^2}^2  +  \| R(t) \|_{\ell^2}^2
+ \| \varepsilon \dot{v}(t) \|_{\ell^2}^2 + \| v(t) \|_{\ell^2}^2  \leq C E(t).
\end{equation}
\end{lemma}

The  essential point in the proof of the approximation result of Theorem \ref{thmain} is that
the fast dynamics of $v$ can be controlled by the $\| \varepsilon \dot{v}(t) \|_{\ell^2}^2$ term in the energy bound
\eqref{energy-control} and in the energy balance equation \eqref{balance-2}. By using the Cauchy-Schwarz inequality
and the estimates \eqref{pepper0}, \eqref{pepper1}, and \eqref{pepper4}, we obtain the following estimate.

\begin{lemma}
\label{lem44}
Assume that $ \Psi \in C^1([0,T_0],\ell^2) $ is a solution of the scalar equation \eqref{Psieq} for some $T_0 > 0$
satisfying (\ref{bound-on-Psi}) for some small $C_0 > 0$. Then there exists constants $ C_1, C_2, C_3 > 0 $ that
depends on $\Psi$ such that for all $ \varepsilon \in (0,1) $ we have
\begin{equation}
\label{energy-balance}
\frac{d}{dt} E(t)  \leq  C_1 E(t)^{1/2} + C_2 E(t) + C_3 \varepsilon E(t)^{3/2}, \quad t \in [0,T_0].
\end{equation}
\end{lemma}

We can now conclude the proof of Theorem \ref{thmain}. Let $S(t) := E(t)^{1/2}$. The initial bound (\ref{initial-bound})
yields $S(0) \leq C_0$ for some $C_0 > 0$ independently of $\varepsilon \in (0,\varepsilon_0)$.
The energy balance estimate (\ref{energy-balance}) can be rewritten in the form
\begin{equation}
\label{energy-balance-S}
\frac{d}{dt} S(t)  \leq  C_1 + C_2 S(t) + C_3 \varepsilon S(t)^2, \quad t \in [0,T_0],
\end{equation}
where the constants $C_1, C_2, C_3 > 0$ have been redefined. Let $T_*$ be defined by
$$
T_* := \sup \{ T > 0 : \quad S(t) \leq \varepsilon^{-1} C_3^{-1} C_2, \quad t \in [0,T] \},
$$
for the given constants $\varepsilon$, $C_2$, and $C_3$. Then, by Gronwall's inequality,
we obtain
$$
S(t) \leq \left[ S(0) + (2 C_2)^{-1} C_1 \right] e^{- 2 C_2 t} \leq \left[ C_0 + (2 C_2)^{-1} C_1 \right] e^{-2 C_2 T_0}, \quad t \in [0,T_0].
$$
Since $T_0 < T_*$ if $\varepsilon > 0$ is appropriately small, we obtain $S(t) \leq C$ for some $C > 0$ independently
of $\varepsilon \in (0,\varepsilon_0)$, so that the final bound (\ref{final-bound}) holds.
The approximation result of Theorem \ref{thmain} is proven.

\section{Generalization}
\label{secdisc}

We have proven the approximation result of Theorem \ref{thmain} for the simplest nonlinear interaction potential
$  W'(u) =  u +  u^2 $. For a more general interaction potential $W \in C^3(\mathbb{R})$, Taylor expansions around $U$ yield
\[
W'(U_{j}-w_{j+1}) + W'(U_{j}+w_{j+1})  
 	 =  2 W'(U_{j}) + \mathcal{O}(|w_{j+1}|^2 )
\]
and
\[
W'(U_{j}-w_{j+1}) - W'(U_{j}+w_{j+1})  
 	 =  - 2 W''(U_{j}) w_{j+1} + \mathcal{O}(|w_{j+1}|^2 ),
\]
so that the system of coupled equations \eqref{eq-U} and \eqref{eq-w} is rewritten in the more general form:
\begin{eqnarray}
\label{eq-U-gen}
\ddot{U}_{j} + W'(U_j) + \mathcal{O}(|w_{j+1}|^2) & = &  g(U_{j+2},U_{j-2},w_{j+3},w_{j-1}), \\
\label{eq-w-gen}
\varepsilon^2 \ddot{w}_{j+1} + (2 + \varepsilon^2) W''(U_j) w_{j+1} + \mathcal{O}(|w_{j+1}|^2)
& = &  \varepsilon^2 h(U_{j+2},U_{j-2},w_{j+3},w_{j-1}),
\end{eqnarray}
The energy function for the perturbation terms in the decomposition \eqref{decomposition} becomes
\begin{eqnarray}
\label{energy-linear-gen}
	E(t) = \frac{1}{2} \sum_{j \in 2 \Z} \dot{R}_j^2 + W''(\Psi_j) R_j^2 + \varepsilon^2 \dot{v}_{j+1}^2 + 2
W''(\Psi_j + R_j) v_{j+1}^2.
\end{eqnarray}
It follows by repeating the previous analysis that the same approximation result stated in Theorem \ref{thmain}
applies to the more general interaction potential satisfying the conditions $W \in C^3(\mathbb{R})$ and $W''(0) > 0$.

\bibliographystyle{plain}
\bibliography{fpubib}

\begin{thebibliography}{10}

\bibitem{BSSZ19}
S.~Baumstark, G.~Schneider, K.~Schratz, and D.~Zimmermann.
\newblock Effective slow dynamics models for a class of dispersive systems.
\newblock {\em J Dyn Diff Equat}, 2019.

\bibitem{BP13}
M.~Betti and D.E. Pelinovsky.
\newblock Periodic travelling waves in diatomic granular chains.
\newblock {\em Journal of Nonlinear Science}, 23:689--730, 2013.

\bibitem{ph1}
M.A. Colin.
\newblock Solitons in the diatomic chain.
\newblock {\em Physical Review A}, 31(3):1754--1762, 1985.

\bibitem{FH20}
T.E. Faver and H.J. Hupkes.
\newblock Micropteron traveling waves in diatomic fermi--pasta--ulam--tsingou
  lattices under the equal mass limit.
\newblock {\em Physica D}, 2020.

\bibitem{FW18}
T.E. Faver and J.D. Wright.
\newblock Exact diatomic fermi--pasta--ulam--tsingou solitary waves with
  optical band ripples at infinity.
\newblock {\em SIAM Journal of Mathematical Analysis}, 50:182--250, 2018.

\bibitem{GMWZ14}
J.~Gaison, S.~Moskow, Wright J.D., and Q.~Zhang.
\newblock Approximation of polyatomic fpu lattices by kdv equations.
\newblock {\em Multiscale Model. Simul.}, 12:953--995, 2014.

\bibitem{GSWW19}
N.~Giardetti, A.~Shapiro, S.~Windle, and Wright J.D.
\newblock Metastability of solitary waves in diatomic fput lattices.
\newblock {\em Mathematics in Engineering}, 1:419--433, 2019.

\bibitem{HW17}
A.~Hoffman and J.D. Wright.
\newblock Nanopteron solutions of diatomic fermi--pasta--ulam--tsingou lattices
  with small mass-ratio.
\newblock {\em Physica D}, 358:33--59, 2017.

\bibitem{ph2}
G.~Huang.
\newblock Soliton excitations in one-dimensional diatomic lattices.
\newblock {\em Physical Review B}, 51(18):12347--12360, 1995.

\bibitem{J12}
G.~James.
\newblock Periodic travelling waves and compactons in granular chains.
\newblock {\em Journal of Nonlinear Science}, 22:813--848, 2012.

\bibitem{JN04}
G.~James and P.~Noble.
\newblock Breathers on diatomic {F}ermi-{P}asta-{U}lam lattices.
\newblock {\em Phys. D}, 196(1-2):124--171, 2004.

\bibitem{JN08}
G.~James and P.~Noble.
\newblock Weak coupling limit and localized oscillations in euclidean invariant
  hamiltonian systems.
\newblock {\em Journal of Nonlinear Science}, 18:433--461, 2008.

\bibitem{chain2}
K.R. Jayaprakash, Yu. Starosvetsky, and A.F. Vakakis.
\newblock New family of solitary waves in granular dimer chains with no
  precompression.
\newblock {\em Physical Review E}, 83:036606 (11 pages), 2011.

\bibitem{chain3}
K.R. Jayaprakash, Yu. Starosvetsky, A.F. Vakakis, and O.V. Gendelman.
\newblock Nonlinear resonances leading to strong pulse attenuation in granular
  dimer chains.
\newblock {\em Journal of Nonlinear Science}, 23(3):363--392, 2013.

\bibitem{LSM97}
R.~Livi, M.~Spicci, and R.~S. MacKay.
\newblock Breathers on a diatomic {FPU} chain.
\newblock {\em Nonlinearity}, 10(6):1421--1434, 1997.

\bibitem{L20}
C.J. Lustri.
\newblock Nanoptera and stokes curves in the 2-periodic
  fermi--pasta--ulam--tsingou equation.
\newblock {\em Physica D}, 402:132239 (13 pages), 2020.

\bibitem{LP18}
C.J. Lustri and M.A. Porter.
\newblock Nanoptera in a period-2 toda chain.
\newblock {\em SIAM Journal of Applied Dynamical Systems}, 17:1182--1212, 2018.

\bibitem{chain1}
M.A. Porter, C.~Daraio, I.~Szelengowics, E.B. Herbold, and P.G. Kevrekidis.
\newblock Highly nonlinear solitary waves in heterogeneous periodic granular
  media.
\newblock {\em Physica D}, 238:666--676, 2009.

\bibitem{Qi14}
W.X. Qin.
\newblock Modulation of uniform motion in diatomic {F}renkel-{K}ontorova model.
\newblock {\em Discrete Contin. Dyn. Syst.}, 34(9):3773--3788, 2014.

\bibitem{Qi15}
W.X. Qin.
\newblock Wave propagation in diatomic lattices.
\newblock {\em SIAM Journal of Mathematical Analysis}, 47(1):477--497, 2015.

\bibitem{ph3}
Y.~Tabata.
\newblock Stable solitary wave in diatomic toda lattice.
\newblock {\em Journal of Physical Society of Japan}, 65(12):3689--3691, 1996.

\bibitem{Vain16}
A.~Vainchtein, Y.~Starosvetsy, J.D. Wright, and R.~Perline.
\newblock Solitary waves in diatomic chains.
\newblock {\em Physical Review E}, 93(4):042210, 2016.

\bibitem{W19}
J.A.D. Wattis.
\newblock Asymptotic approximations to travelling waves in the diatomic
  fermi--pasta--ulam lattice.
\newblock {\em Mathematics in Engineering}, 1:327--342, 2019.

\bibitem{Yoshimura11}
K.~Yoshimura.
\newblock Existence and stability of discrete breathers in diatomic
  fermi--pasta--ulam type lattices.
\newblock {\em Nonlinearity}, 24(4):293--317, 2011.

\end{thebibliography}

\end{document}